\documentclass[12pt]{article}
\usepackage{graphicx}
\usepackage{natbib}
\usepackage{authblk}

\begin{document}
\title{Quantifying the uncertainty of variance partitioning estimates of ecological datasets} 
\author[1]{Matthias M. Fischer}
\affil[1]{\small Freie Universit\"at Berlin, Institut f\"ur Biologie, Mikrobiologie, 14195 Berlin, Germany. Tel.: +49 30 838 53373, E-Mail: \texttt{m.m.fischer@fu-berlin.de}}
\maketitle

\setlength{\parindent}{1cm}
  
\begin{abstract}
An important objective of experimental biology is the quantification of the relationship between predictor and response variables, a statistical analysis often termed variance partitioning (VP). In this paper, a series of simulations is presented, aiming to generate quantitative estimates of the expected statistical uncertainty of VP analyses. We demonstrate scenarios with considerable uncertainty in VP estimates, such that it significantly reduces the statistical reliability of the obtained results. Especially when a predictor variable of a dataset shows a low between-group variance, VP estimates may show a high margin of error. This becomes particularly important when the respective predictor variable only explains a small fraction of the overall variance, or the number of replicates is particularly small. Moreover, it is demonstrated that the expected error of VP estimates of a dataset can be approximated by bootstrap resampling, giving researchers a tool for the quantification of the uncertainty associated with an arbitrary VP analysis. The applicability of this method is demonstrated by a re-analysis of the Oribatid mite dataset introduced by Borcard and Legendre in 1994 and the Barro Colorado Island tree count dataset by Condit and colleagues. We believe that this study may encourage biologists to approach routine statistical analyses such as VP more critically, and report the error associated with them more frequently.
\end{abstract}


\section{Introduction}
Many studies in experimental biology address how a set of variables, commonly termed the predictor variables, relate to a variable or to a set of variables of interest, called response variables. Often, a quantification of the strength of such relationships is desired, in which case a statistical analysis known as variance partitioning (VP) can be used \citep{vp}. Although variance partitioning represents a general purpose statistical tool widely applied throughout biology \citep{genex1, genex1.5, genex2, genex3}, some specific disciplines such as modern community ecology have made more use of it than others \citep{comex1, comex2,comex3,comex4,comex5,comex6,comex6.5,comex7,comex8,comex9,vp_with_error,comex10,comex11,comex12,comex13,comex13.5,comex14}. A comprehensive review of the importance of the role and general usage of VP techniques in modern community ecology is given by \textit{Dray} and colleagues \citep{comex_review}. \\

VP analyses are often applied to community tables (or ``site-by-species tables''), which represent a straightforward way to summarize ecological datasets and are easy to integrate to existing data such as environmental and spatial data (for example see \citet{community_tables}). The tables contain either abundance or presence-absence data, in which columns represent species and rows represent sampled sites. By applying VP to such community tables, a community ecologist can then identify and quantify the strength of relationships between species occurrence and biotic, abiotic, and spatial variables of the sampled sites. \\

The outcome of variance partitioning is a set of estimates on how much variance in species occurrence is explained by a single predictor or through combinations. Because estimates represent averages based on the entire dataset, they can be expected to contain a relatively low uncertainty, $i.e.$ error. Under specific conditions, however, this error could account for a large fraction of the observed variability and significantly skew the results. Apart from measurement errors committed during collection of the primary data and the finite resolution of the used measuring devices, uncertainty in VP estimates is mainly caused by the impossibility to assay all individuals of an ecological community. As a result, sampling effects occur, which may vary considerably in severity. Stochasticity in environmental processes in the form of random fluctuations in species occurrences cause further uncertainties in the results of the statistical analysis. In literature, it is relatively uncommon for authors to couple estimates of variance partitioning with metrics of their uncertainty. However, given the technique is so widely used across biology, identifying instances when VP estimates may contain a high margin of error should be helpful in an analysis.\\

Hence, a series of simulations is presented in this work, which aim to quantify the uncertainty of statistical estimates obtained from a typical variance partitioning approach, based on a redundancy analysis. Our datasets are generated by a simplified toy model for the ecology of two species. The results are then correlated with a set of parameters which potentially affect the VP uncertainty. In this way, critical factors influencing the accuracy of VP estimates are identified and strategies for increasing the statistical reliability of VP techniques are given. Moreover, it will be demonstrated that a simple bootstrap resampling procedure as introduced by \textit{Efron \& Tibshiranni} \citep{bootstrap1, bootstrap2} can be used to approximate the expected error of these VP analyses. We also show that in case of more complex datasets closer to ecological reality, bootstrapping still yields very accurate estimates of the expected error of VP analyses, thereby giving researchers a practical and easy to use tool for the quantification of the uncertainty associated with an arbitrary VP analysis. The method is finally used to re-examine the Oribatid mite dataset by \textit{Borcard \& Legendre} \citep{mite} and the Barro Colorado Island tree count dataset by \textit{Condit} and colleagues \citep{BCI} in order to demonstrate its practical applicability.

\section{Materials and Methods}
\subsection{Two-species toy model}
In order to obtain appropriate datasets, we employ a simple toy model for two species sampled at $n = 100$ sites. These sites can be thought of as different physical locations with different abiotic properties. The occurrence of a species at a specific site depends on two numeric abiotic environmental factors inherent to the respective site. These factors, termed $x$ and $y$, are uniformly and randomly drawn from the interval $[0,1]$. The abundances of the two species are then calculated as a function of $x, y$ \citep{hutchinson}. Following the model by \textit{Minchin}, a unimodal relationship between the abundance of a species and each of the environmental factors is assumed, so that each species $i$ possesses an optimum point $(x^*_i|y^*_i)$ at which its abundance reaches its peak. However, in contrast to the generalised beta function introduced by \textit{Minchin} and \textit{Austin}, we assume a simple bell-shape for the species response curve that maps the value of an environmental predictor variable onto the associated abundance of a species. This allows us to approximate the response curves through simple normal distributions \citep{minchin1987simulation, austin1976non}. The relative abundance of the $i$-th species $\alpha_i$ then follows as

\begin{equation}
\alpha_i(x, y) = (f(x, x^*_i, \sigma^2) + \epsilon_{x,i}) \; (f(y, y^*_i, \sigma^2) + \epsilon_{y,i}) ,
\end{equation}

where $f$ denotes the probability density function of the normal distribution with variance $\sigma^2$ as a proxy for the size of the simulated ecological niche. In the following $\sigma$ is set to 0.5 denoting a niche of intermediate size. In order to account for the inherent stochasticity of biological processes, normally distributes noise $\epsilon_{x,i}, \epsilon_{y,i}$ with zero mean (Gaussian white-noise) is added to each predictor before multiplication. A representative plot of such a resulting three-dimensional fitness landscape is given in Figure 1. \\

\begin{figure}[!h]
	\centering
	\includegraphics[width=12cm]{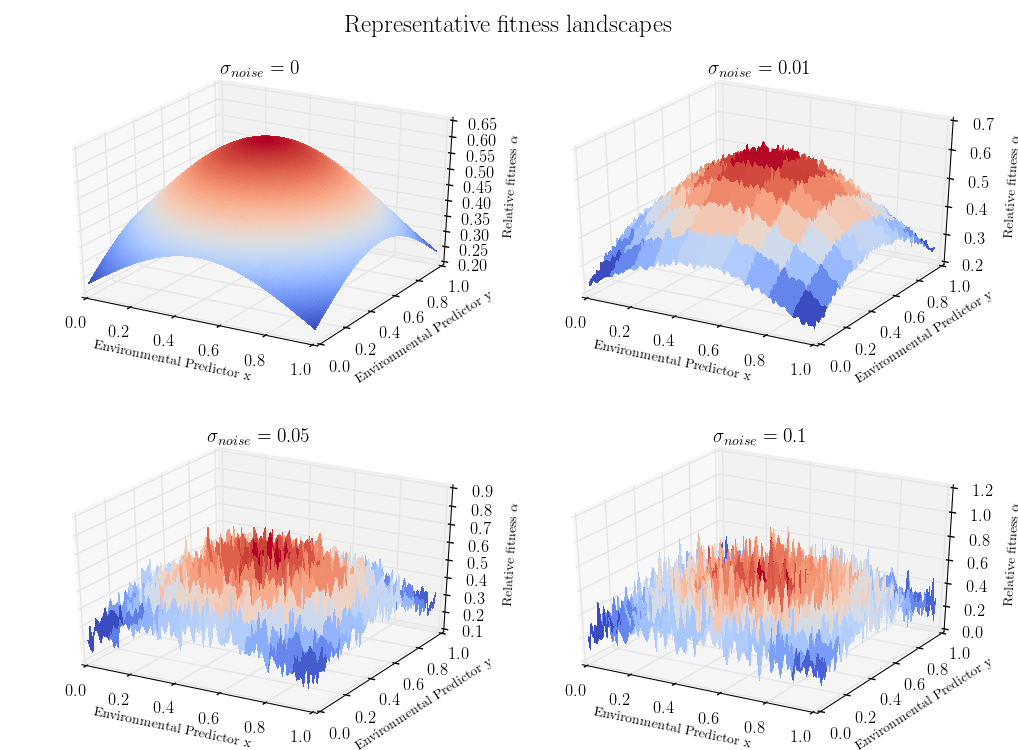}
	\caption{\textit{Representative fitness landscapes simulated with the two species toy model. The optimum of the species response curve lies at $x^* = y^* = 0.5$ and Gaussian noise with a mean of zero and standard deviations of $\sigma_{noise} = 0, 0.01, 0.05, 0.1$ has been added to both predictors before multiplication.}}
\end{figure}

In order to obtain the final population of species $i$, $N_{i}$, the relative abundance $\alpha_i$ is normalized and multiplied by the carrying capacity factor $K=10^{4}$, giving

\begin{equation}
N_i(x,y) =\left[\frac{\alpha_i(x,y)}{\sum_j \alpha_j(x,y)} K \right]
\end{equation}

The result is rounded off to the next positive integer value.

\subsection{Variance partitioning}
In order to assess the influence of factors onto the uncertainty of VP estimates, site-by-species tables are generated with different parameter configurations. These parameters include the number of sampled sites $n$ in the site-by-species table and the scatter range of an environmental predictor variable across the different simulated sites. The latter of them is manipulated by reducing the scatter range of an environmental predictor from $[0; 1]$ to $[0; y_{max}]$ with $y_{max} = 0.1, 0.2, ..., 1.0$. The optimum points of the simulated ecological niches are varied as well by changing the distance between the means of the species response curves. To achieve this, $y^*_1$ of the considered environmental predictor is permanently set to zero, while $y^*_2$ is systematically varied within the interval $[0;1]$. In this way, the relative amount of influence the simulated environmental predictor $y$ exerts on the species abundances dataset can be manipulated: with increasing differences in niche midpoints the influence of this specific abiotic environmental factor increases. In contrast, the scatter range of the predictor $x$ remains constant at $[0; 1]$, as well as the associated optimum points of the species response curves $x^*_1, x^*_2$. Finally, we also vary the standard deviation $\sigma_{noise}$ of the normally distributed noise $\epsilon_{y,i}$. All scenarios were analysed for standard deviations of $\sigma_{noise} = 0.01$, $\sigma_{noise} = 0.05$ and $\sigma_{noise} = 0.1$. \\

Each individual scenario is simulated $M = 1000$ times. Every generated site-by-species table is subsequently subjected to the VP analysis provided in the 'vegan' R package, which is based on redundancy analysis ordination (RDA) \citep{vegan}. RDA uses linear regression of multiple response variables on multiple predictor variables. The results of these regressions are then subject to principal components analysis. In this way, linear relationships between components of the response variables "redundant with" ($i.e.$ explained by) a set of predictors can be detected. \\
All $M$ VP estimates, expressed as the adjusted $R^2$ of the considered environmental predictor, are recorded and their standard deviation is calculated as a proxy of the associated absolute uncertainty. These uncertainties can then be correlated with the parameter configurations of the individual simulated scenarios in order to analyse the influence of these parameters on the uncertainty of the VP estimates.

\subsection{Bootstrap resampling}
For each scenario analysed, ten additional site-by-species tables are generated and subjected to a bootstrap resampling procedure as laid out by \textit{Efron \& Tibshirani} \citep{bootstrap1, bootstrap2}. From each dataset the rows are resampled randomly with replacement $M$ times, in other words the dataset is bootstrapped across sites. For all $M$ subsamples, a variance partitioning is performed and all VP estimates are recorded and their standard deviation is calculated as a metric of absolute uncertainty. The relative uncertainty is then calculated by dividing the standard deviation by the mean $R^2$ value of the respective simulation run. The ten relative uncertainties of a scenarios are then averaged and correlated with the relative uncertainty estimates obtained from the previous method.

\subsection{Validation with more realistic synthetic datasets}
In order to show that bootstrap resampling can also be used to obtain an estimate of the expected error in case of more complex, and thereby more ecologically realistic datasets, the data generation model described before was adjusted: In contrast to the previous approach, the abundance of five species is simulated while we vary their optimum points as well. Specifically, the latter are uniformly and randomly drawn from the interval $[0, 1]$. The resulting abundance curves can then be expected to show a strong amount of nonlinearity \citep{minchin1987simulation}. This implies that the previous VP approach, based on redundancy analysis, cannot be used as it assumes a linear relationship between predictors and response variables. Instead, we apply canonical correspondence analysis (CCA), a multivariate statistical technique introduced by \textit{ter Braak} \citep{cca} which is able to deal with both linear and nonlinear relationships between predictor and response variables \citep{rda_vs_cca}. Following the standard procedure, the generated abundance data are log-transformed before the actual CCA in order to limit the influence of outlier values on the fitted models \citep{log}. \\
It should be noted that CCA does not analyse the influence of the isolated predictor variables, but instead computes a set of linear combinations of the predictors as ordination axes \citep{ordination_axes}. Therefore we assessed the uncertainty in the amount of the total variance explained by all ordination axes together which was subsequently compared to the overall variance of the complete dataset. This was done for sample sizes of $n = 20, 40, ... 100$ simulated sites and normally distributed noise added to both predictor variables with a standard deviation of $\sigma_{noise} = 0, 0.01, 0.05, 0.1$. This yielded a total of twenty different scenarios. Each scenario was repeated five times with different, uniformly and randomly chosen mean values for the response curves of the simulated five species. 

\subsection{Application to a real-world dataset}
In order to practically test the bootstrap method as a means of quantifying VP error, we re-analysed the Oribatid mite dataset by {\it Borcard \& Legendre} \citep{mite} using Canonical Correspondence Analysis. We performed $M = 1000$ bootstrap resamples of the original dataset and quantified the mean and standard deviation of both the variance explained by the environmental and the spatial predictors. We repeated this analysis ten times in order to check, if the analysis yields reproducable results.

\section{Results}
\subsection{Drivers of uncertainty in the two-species toy-model}
\subsubsection{Sample size}
A strong negative effect of the sample size $n$ on the uncertainty of the associated VP estimates could be found as demonstrated by Figure 2. For minimal and intermediate amounts of noise, the relative error of the VP estimates sharply decreases with increasing $n$ and becomes barely noticeable for a high sample size of $n=1000$. However, for high amounts of noise (dashed green line), a substantial amount of error remains even for high sample sizes.

\begin{figure}[!h]
	\centering
	\includegraphics[width=10cm]{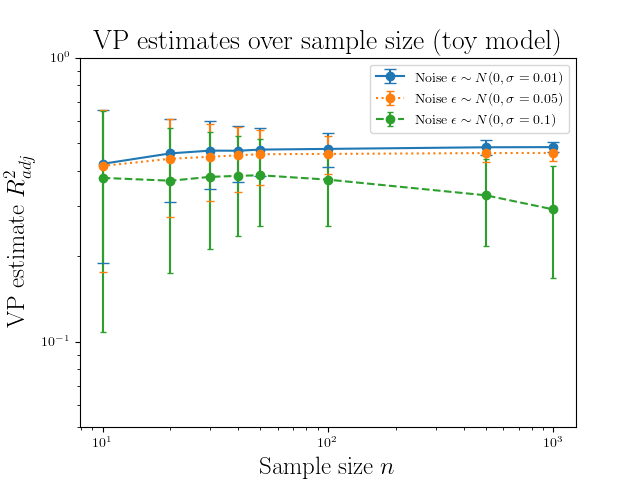}
	\caption{\textit{Effect of the sampling size $n$ on adjusted $R^2$ obtained by VP analysis. Error bars represent the standard error of the mean across all simulated replicates. Note the log-scaling of the y-axis and the increase in relative uncertainty for smaller sample sizes.}}
\end{figure}

\subsubsection{Sampling range of a predictor}
A strong negative effect of sampling range of a predictor variable on the uncertainty of its VP estimates could be demonstrated as shown in Figure 3. The amount of noise added to a predictor variable positively influences the uncertainty of the adjusted $R^2$ value associated with the respective predictor. However, for sufficiently wide sampling ranges the influence from the amount of added noise strongly decreases. Also note the strong negative influence of narrow sampling ranges on the adjusted $R^2$ values, demonstrating that a sufficiently small sampling range can lead to a strong underestimation of the influence a predictor exerts on the measured response variables.

\begin{figure}[!h]
	\centering
	\includegraphics[width=10cm]{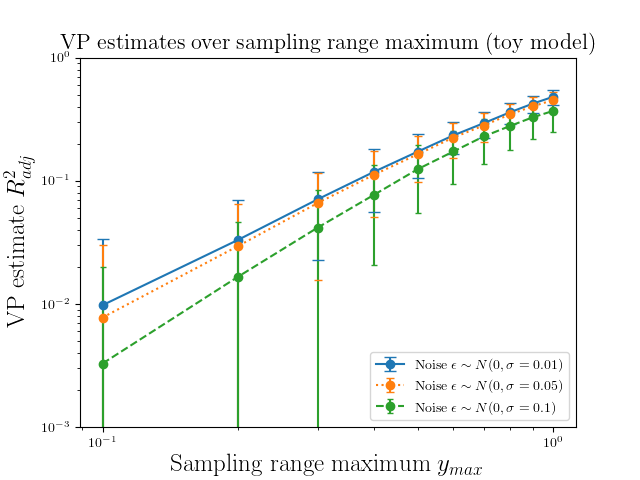}
	\caption{\textit{Influence of the sampling range width on adjusted $R^2$ obtained by VP analysis. Error bars represent the standard error of the mean across all simulated replicates. Note the log-scaling of the y-axis and the increase in relative uncertainty for smaller sample range maxima, $i.e.$ for smaller sampling ranges.}}
\end{figure}

\subsubsection{Effect size of a predictor}
Similarly, a strong negative effect of the difference between optimal values for the two species could be demonstrated, as shown in Figure 4. In other words: the bigger the difference in the optima of the two species response curves is, the more precisely the influence of this predictor will be estimated. The amount of noise added to a predictor variable again positively influences the uncertainty of the adjusted $R^2$ values. However, for sufficiently large optimum differences, the influence on VP uncertainty strongly decreases.

\begin{figure}[!h]
	\centering
	\includegraphics[width=10cm]{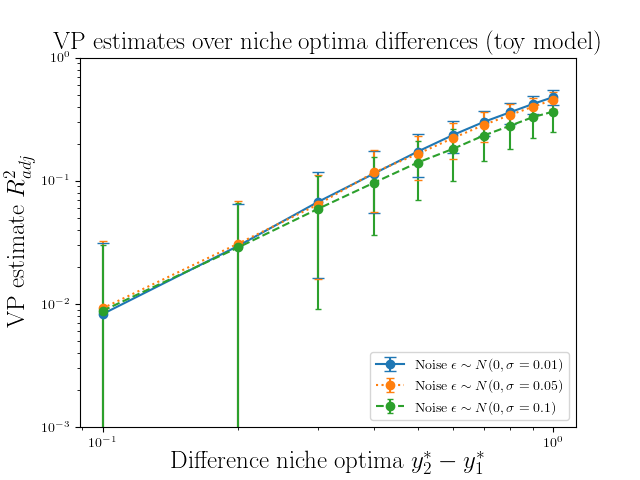}
	\caption{\textit{Effects of the difference in optimal values in species response curves on adjusted $R^2$ obtained by VP analysis.  Error bars represent the standard error of the mean across all simulated replicates. Note the log-scaling of the y-axis and the increase in relative uncertainty for smaller differences in niche optima of the species, $i.e.$ for a smaller predictive power of an environmental predictor variable.}}
\end{figure}

\subsection{Bootstrap estimation for the two-species toy model}
A highly significant positive linear relationship between the observed relative errors of the simulated datasets and the uncertainties estimated by bootstrapping could be observed (Figure 5). In the range of 3 \% to 100 \% relative error, the two samples show a strong correlation of approximately $0.93$ (Pearson's $r = 0.931$, $t = 20.901$, $df = 67$, $p < 0.001$***). For relative errors above 100 \% however, the bootstrapping method significantly over-estimates the expected error of VP estimates.

\begin{figure}[!h]
	\centering
	\includegraphics[width=10cm]{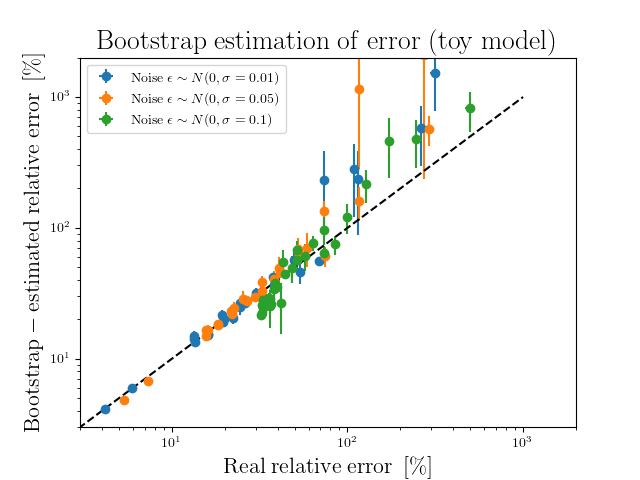}
	\caption{\textit{Correlation of the observed relative error and the bootstrap-estimated error. Notice the strong correlation for relative uncertainties up to 100 \%. Dashed line represents the diagonal $y=x$.}}
\end{figure}

\subsection{Bootstrap estimation for a more complex dataset}
Again, a highly significant positive linear relationship between the observed relative errors of the simulated datasets and the uncertainties estimated by bootstrapping could be observed (Figure 6). The two samples show a strong correlation of approximately $0.98$ (Pearson's $r = 0.981$, $t = 49.38$, $df = 98$, $p < 0.001$***). However, it should be noted that this approach only yielded errors in the range between $0.5 \%$ and $\sim 30 \%.$ Even with a doubled standard deviation of the noise of $\sigma_{noise} = 0.2$, higher margins of error were not achievable.

\begin{figure}[!h]
	\centering
	\includegraphics[width=10cm]{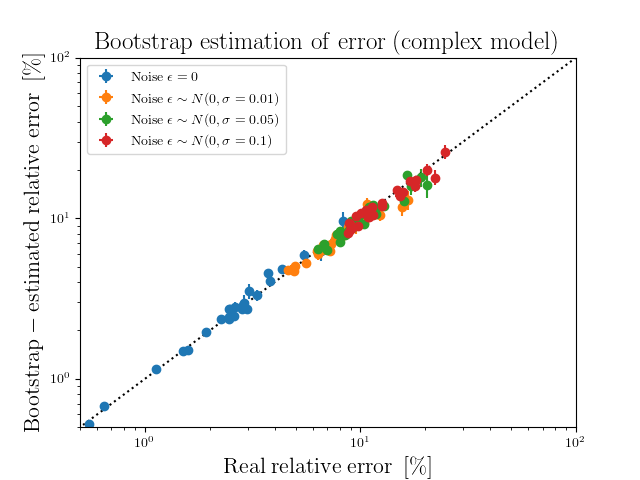}
	\caption{\textit{Correlation of the observed relative error and the bootstrap-estimated error. Notice the consistently good approximation of the expected error. Dashed line represents the diagonal $y=x$.}}
\end{figure}

\subsection{Application to real-world datasets}
Approximately 23\% of the overall variance of the Oribatid mite dataset was explained by purely environmental predictors (0.95-CI: [18.0\%; 30.7\%]) with a standard deviation of approximately 3.3\%, equalling a relative uncertainty of approximately 14\%. Additional 37\% of the variance (0.95-CI: [28.7\%; 45.0\%]) were explained by spatial variables with a standard deviation of approximately 4\%, corresponding to a relative uncertainty of 12\%. The remaining 40\% of the variance of the dataset (0.95-CI: [32.4\%; 47.2\%], SD: 3.7\%, equalling a relative uncertainty of approximately 9\%) are not explained by either of the two predictors.  \\
In case of the Barro Colorado Island tree count dataset, approximately 16.7\% (0.95-CI: [14.4\%; 19.6\%], SD: 1.33 \%, relative uncertainty 8\%) of species abundance could be explained by environmental factors, additional 31.2\% (0.95-CI: [22.7\%; 40.4\%], SD: 4.57 \%, relative uncertainty 14.6\%) by spatial variables. The remaining 52.2\% (0.95-CI: [42.4\%; 60.3\%], SD: 4.66 \%, relative uncertainty 8.99\%) of the variance could not be attributed to either of these factors.

Across all ten repetitions of both analyses, no significant differences could be detected. Figure 7 shows the results of respectively one representative analysis for both of the datasets. Every repetition took within fourteen to fifteen seconds to complete on a 64 bit notebook with a 2.3 GHz AMD double core processor running Ubuntu Linux 14.04 LTS.

\begin{figure}[!h]
	\centering
	\includegraphics[width=10cm]{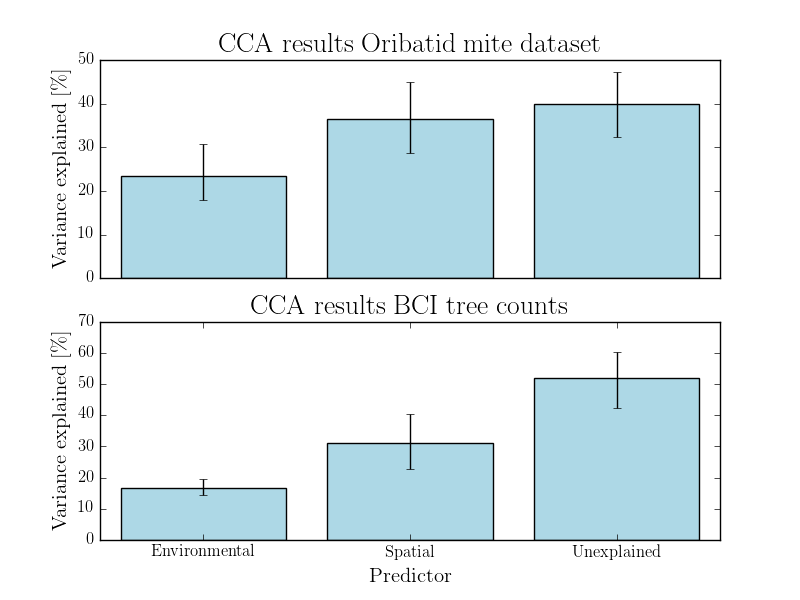}
	\caption{\textit{Results of one representative bootstrap analysis of the Oribatid mite dataset and the Barro Colorado Island tree count dataset respectively. Error bars represent bootstrapped 95\% confidence interval of adjusted $R^2$ values.}}
\end{figure}

\section{Discussion}
In the previous section, the influence of three main parameters on the uncertainty of VP estimates was assessed. The results are in line with the already existing literature on the reliability of statistical models. It is a well established fact that the sample size crucially influences the precision with which an unknown parameter can be estimated in an empirical study as described by the law of large numbers. Therefore, the clear inverse relationship between sample size, in this case the number of row entries in the sampled site-by-species table, and the uncertainty of VP estimates demonstrated in this study is not a surprising result. The negative relationship between the sampling range of a predictor variable and the uncertainty of its associated VP estimate is in accordance with existing research as well. As pointed out by $e.g.$ \textit{Hill \& Tiedeman} for the case of linear and nonlinear regression techniques, an increasing sampling range will lead to a higher amount of noise which a regression model can tolerate while still being able to accurately estimate the respective model parameters \citep{range}. Analogously, in case of a fixed amount of random noise in a dataset, a statistical model can be expected to operate with higher precision as the sampled range of a predictor increases. Finally, the similar inverse relationship between the effect size of a predictor (in this paper expressed as the difference in the optimum values of the two species for an environmental predictor) and the uncertainty of the associated VP estimates was to be expected as well. Clearly, a higher effect size compared to the added random noise will lead to a more precise estimate of the relative importance of the respective predictor variable, $i.e.$ a lower uncertainty of the associated VP estimate.\\

Based on these results, general ways to reduce the uncertainty of VP estimates and their interpretation for future experimental studies can be proposed: The results of the simulations demonstrate the positive effect of a higher number of observations on the certainty of VP estimates. In case of datasets containing a comparatively small number of observations the results of VP analyses therefore need to be interpreted with a higher degree of caution. With regards to environmental factors that show a lower amount of between-site variation, it is advisable to interpret their VP estimates with caution as well, given their VP estimates are subjected to a significantly higher amount of relative error. Furthermore, such a small sampling range also increases the risk of under-estimating the relative influence of a predictor on the variance of a dataset. However, based on the analyses presented, predictors that show a high amount of variation between the different sites are more suitable for VP analysis, since the relative errors of their associated VP estimates can be expected to remain rather small. Finally, for the interpretation of VP results, focussing on predictors that explain a high amount of the variance in the dependent variables by having a comparatively big effect size, seems advisable. The VP estimates of predictors explaining only a very small share of the variance of a dataset, however, are affected by a higher margin of error, which can significantly reduce their statistical reliability.

When dealing with a real dataset obtained experimentally, the previous findings allow us to only estimate the uncertainty of VP estimates in a purely qualitative way. In order to also obtain quantitative estimates, the bootstrap analysis should be used, as it yields a consistently good and unbiased approximation of the expected error the VP estimates are subjected to. This holds true both in case of the simplified two-species toy model and also in case of more realistic datasets as shown in Figures 5 and 6.

\section{Conclusion}

In this study, we analysed the impact of model parameters on VP uncertainty estimates and found a significant influence of the overall sample size, the sampling range of a predictor and the predictive power of a predictor.  We demonstrated that the expected errors of VP estimates can be approximated reasonably well by performing a simple bootstrap resampling \citep{bootstrap1, bootstrap2}, both in case of the simplified toy model and also with more complex datasets closer to ecological reality. We confirmed the feasibility of the method by re-analysing the Oribatid mite dataset by \textit{Borcard \& Legendre} \citep{mite} and demonstrated consistent results. Therefore, bootstrapping seems to be a suitable technique for estimating the amount of error one has to expect when performing VP analyses. For further experimental studies, we suggest including uncertainty estimates obtained by such methods in order to make the presented VP estimates easier to interpret and in order to clarify the fact that they are statistical estimates, which are subjected to a certain amount of uncertainty. \\

Overall it has become clear that a reasonable use of VP techniques requires an awareness of their limitations and reliability. As pointed out by \textit{Gilbert \& Bennett}, the correct interpretation of statistical results is important to both theoretical and applied ecology, especially with regard to the understanding of the theoretical underpinnings of meta-communities and the impact of environmental changes on biodiversity \citep{comex12, vp_importance,vp_numbers_add_up}. We believe that the findings presented in this work may encourage biologists to approach routine statistical analyses such as VP more critically and report associated error more frequently. We furthermore hope to have encouraged further research into the limitations and reliability of statistical techniques, in order to increase their usefulness for future experimental studies. \\

\section{Acknowledgements}
I would like to thank M.Sc. Joscha Reichert for his extensive and extremely helpful comments on an earlier version of the manuscript. I also would like to thank Dr Stavros D. Veresoglou for providing the initial research question and his guidance and assistance throughout this research project.

\footnotesize
\bibliography{references} 
\bibliographystyle{apalike}

\end{document}